\documentclass{optica-article}

\journal{opticajournal} 

\articletype{Research Article}

\usepackage{lineno}

\newcommand{\FT}{\mathcal{F}}
\newcommand{\IFT}{\mathcal{F}^{-1}}
\newcommand{\fnum}{\mathfrak{f}}
\newcommand{\Fprop}{\mathcal{D}}
\newcommand{\lnorm}[1]{\big\|{#1}\big\|}
\newcommand{\inv}{^{-1}}

\newcommand{\RR}{\mathbb{R}}

\DeclareMathOperator*{\argmin}{arg\,min}

\DeclareMathOperator{\eye}{\mathbb{I}}

\DeclareMathOperator{\diag}{diag}

\newcommand{\mynote}[1]{\textcolor{red}{#1}}
\renewcommand{\mynote}[1]{}

\usepackage{graphicx}

\usepackage{siunitx}

\begin{document}

\title{Phase retrieval beyond the homogeneous object assumption for X-ray in-line holographic 
imaging}

\author{Jens Lucht,\authormark{1,2,*} Leon M. Lohse,\authormark{1,2,3} Thorsten
Hohage,\authormark{4} and Tim Salditt\authormark{1,2}}

\address{\authormark{1}Georg-August-Universität Göttingen, Institut für Röntgenphysik,
Friedrich-Hund-Platz 1, D-37077 Göttingen, Germany\\
\authormark{2}Max Planck School of Photonics,  Albert-Einstein-Straße 15, D-07745 Jena, Germany\\
\authormark{3}Deutsches Elektronen-Synchrotron DESY,
Notkestraße 85, D-22607 Hamburg, Germany\\
\authormark{4}Georg-August-Universität, Institut für Numerische und Angewandte Mathematik
Göttingen, Lotzestraße 16-18, D-37083 Göttingen, Germany}

\email{\authormark{*}jens.lucht@uni-goettingen.de} 


\begin{abstract*}
X-ray near field holography has proven to be a powerful 2D and 3D imaging technique with applications ranging from biomedical research to material sciences.
To reconstruct meaningful and quantitative images from the measurement intensities, however, it relies on computational phase retrieval which in many cases assumes the phase-shift and attenuation coefficient of the sample to be proportional.
Here, we demonstrate an efficient phase retrieval algorithm that does not rely on this homogeneous-object assumption and is a generalization of the well-established contrast-transfer-function (CTF) approach.
We then investigate its stability and present an experimental study comparing the proposed algorithm with established methods.
The algorithm shows superior reconstruction quality compared to the established CTF-based method at similar computational cost.
Our analysis provides a deeper fundamental understanding of the homogeneous object assumption and the proposed algorithm will help improve the image quality for near-field holography in biomedical applications.
\end{abstract*}

\section{Introduction}

X-ray phase contrast imaging tomography (XPCT) enables three-dimensional (3D) imaging at scalable field of view and resolution for a wide range of applications from biomedical research to material sciences \cite{Salditt__2020,Withers_NatRev_2021}.
Due to the penetration power of hard X-rays, XPCT is compatible with bulk specimen, complex sample environments, and in operando configurations.  
For samples which lack sufficient absorption such as soft unstained tissues, or more generally soft matter, 
phase contrast based on free propagation and self-interference has significantly extended the applicability and use of computed tomography (CT). Compared to other
variants of X-ray phase contrast imaging (XPCI) such as grating-based, speckle-based, or ptychographic approaches, propagation based imaging (PBI) is particularly attractive, due to the lack of additional optical elements, and the fact that no lateral scanning of the sample is required.
Combined with synchrotron radiation of high spatial coherence, resolution in the sub-micrometer regime is routinely available, and even sub-\qty{100}{\nm} can be achieved when using divergent beams and geometric magnification \cite{Salditt_JoSR_2015,Morawe2015}. For two-dimensional (2D) imaging (XPCI), full field imaging with sub-\qty{15}{\nm} has been recently demonstrated \cite{Soltau:21}.

A crucial step for XPCI and XPCT is the phase retrieval step, where the task is to recover the phase
from the intensity-only measurements, see for example \cite{MaretzkeHohage_NanoscalePhotonicImaging_2020}. 
In PBI, the signal is given by near field 
diffraction patterns or inline \textit{holograms}, formed by self interference downstream from the sample as governed by the Fresnel number $\fnum$. These interference patterns are the input of the phase retrieval and reflect the local phase shifts which the object has imparted upstream on the beam. Note, that parallel beam or cone beam illuminations are equally possible and  equivalent up to geometric magnification and an effective rescaling of the propagation distance, 
as expressed by the Fresnel scaling theorem \cite{Paganin2006}. In fact, 
in the experimental part of this work, we rely on a waveguide, acting as quasi point source and generating an aberration free divergent wavefront for magnification.

Phase retrieval for PBI is actively researched and numerous algorithms have been developed so far. Examples are transport-of-intensity (TIE) based approaches \cite{Paganin2002,Witte2009}, which are restricted to large $\fnum$ and are often for data acquired at laboratory X-ray sources,
iterative projection type algorithms like Gerchberg-Saxton and its derivatives\cite{Gerchberg1972,Luke2004,Hagemann2018}, as well as gradient decent-type algorithms \cite{Davidoiu2011,Wittwer2022,Huhn2022}. In particular for CT, where easily up a few thousands of projections
are acquired, computational efficiency is very important, and the vast majority of XPCT data is phased
by direct but approximative methods, notably in the TIE regime by Paganin's method \cite{Paganin2002}, and in at small $\fnum$ by the contrast transfer function (\textit{CTF}) approach \cite{Cloetens1999}. Hence, phase
retrieval is mostly performed by regularized Fourier filtering. This is fast for sure, but is it also sufficiently exact and general enough for the wide range of applications?
In the derivations of the most common phase retrieval schemes, the assumption of optically weak and/or homogeneous objects is ubiquitous and presumingly overly restrictive. Strictly speaking, this results in an inexact model and inconsistency with the acquired data. It may be expected, that these inconsistencies cause
artifacts in reconstructions, which degrade reconstruction quality. In the present work, we
generalize CTF-based phase retrieval to independent (uncoupled) reconstruction of phase and amplitude, 
with quantified stability analysis, and present a detailed analysis of the homogeneity assumption within
the CTF-framework. We study the consequences of posing or relaxing the assumptions for different parameter settings based on numerical simulations and experimental data acquired for a non-homogeneous test object.


Closely related to the presented approach is \cite{Kostenko2013}, where the authors compare
frequency-dependent Tikhonov and total variation (TV) regularization within the CTF approximation.
In contrast to our approach constraints on the object, such as (non-)negativity or compact support,
cannot be imposed on the solution. Moreover, a different iteration scheme is used.
In \cite{Villanueva2017} the CTF is combined with Alternating Direction Method of Multipliers (ADMM),
but under the assumption of pure-phase or single material objects  and the use of sparsity-promoting regularization. Possibly, \cite{Mom2022} is the closest related
work to the present one, where the authors present a CTF-based single-distance reconstruction without homogeneity assumption but with a TV-prior. However, in the present work, we focus on a comparison and analysis of the homogeneity assumption within
the CTF-framework while not relying on TV priors.

The manuscript is organized as follows: After this introduction, section 2 presents the basic theory, starting with the conventional CTF which we rephrase for notational clarity before giving the generalisation
for the non-homogeneous case. In section 3, the results of the stability analysis are given for the relevant experimental parameters, and the corresponding regularization strategy as a function of the given number of different defocus distances. Finally, 
state-of-the-art nanoscale XPCI experiments are shown in section 4, and are used to test and illustrate the 
capability of the present phase retrieval approach.

\section{Fresnel near-field holography}
Let $F$ be our Fresnel imaging (forward) operator given by
\begin{equation}
 I \approx F(f) \equiv \left| \Fprop\left(\exp(f)\right) \right|^2
 \label{eq:fresnel_imag}
\end{equation}
with $I$ the measured (noisy) intensities $I$, 
the object to reconstruct $f = \mu + i\phi$ under the projection approximation, given by a phase shift 
$\phi\leq 0$ and absorption $\mu\leq 0$, and
the Fresnel propagator $\Fprop$ for any suitable function $\psi$ given by~\cite{Goodman2007}
\begin{equation}\label{eq:Fresnel_prop}
 \Fprop(\psi) \equiv \IFT\left( m_\fnum(\xi) \cdot \FT(\psi)(\xi) \right),
\end{equation}
with transfer function
$
 m_\fnum(\xi) = \exp\left( -\frac{i|\xi|^2}{4\pi\fnum} \right),
$
two-dimensional Fourier transform $\FT$, respectively its inverse $\IFT$, and spatial frequency $\xi$.
We call the intensity images \textit{holograms} due to their in-line
holographic origin. We see, $F$ clearly is a nonlinear operator due to the complex exponential
and squared modulus. Experimental parameters can be condensed into a dimensionless quantity known
as Fresnel number, given by
$
 \fnum = \frac{s^2}{\lambda z_{12}},
$
with photon wavelength $\lambda$, free-space propagation distance $z_{12}$ between object and
detector,
and reference scale $s$. We set this scale to the pixel size of the imaging system $\Delta_x$ and
will use the term Fresnel number for this \textit{pixel Fresnel
number}.


In projection approximation the relation between the energy $E$ and 3D-spatially $(x,y,z)$ dependent
(complex) refractive index $n(E; x,y,z) = 1 - \delta(E; x,y,z) + i\beta(E; x,y,z)$ and measured
\textit{2D-projection images} is given by,
$
  \phi(E;x,y) = -k \int \delta(E;x, y, z) \,\mathrm{d}z
$
and 
$
  \mu(E;x,y)= -k \int \beta(E; x, y, z)\, \mathrm{d}z,
$
respectively, where the optical axis in parallel beam geometry is chosen as $z$-axis, and $k = 2\pi/\lambda$ 
is the wave number in vacuum. Henceforward, we
assume a constant energy $E$ and do not state its dependence explicitly. Note that in our
definition phase delay and absorption are non-positive.

An object is called \textit{optically weak} 
if $|\phi|,|\mu|\ll 1$.
For such objects the Fresnel imaging operator \eqref{eq:fresnel_imag} can be approximated by the well-known
\textit{contrast transfer function} (CTF).
Mathematically, the CTF corresponds to linearzation of \eqref{eq:fresnel_imag} with respect to the
object $f$.
The linearized CTF forward operator is given by $L(f)$, where
\begin{equation}
 I \approx L(f) \equiv 1 + 2\Re[\Fprop(f)],
\end{equation}
with real-part operator $\Re$.
More commonly, $L$ is expressed in Fourier space for phase $\phi$ and absorption $\mu$, 
\begin{equation}
\begin{aligned}
 L(f)
& = 1 + 2\IFT\left(
  s_\fnum(\xi) \cdot \FT(\phi)(\xi) +
  c_\fnum(\xi) \cdot \FT(\mu)(\xi)
 \right),\\
&\text{with}
\quad s_\fnum(\xi) \equiv \sin\left(\frac{|\xi|^2}{4\pi\fnum}\right) \quad \text{and} \quad 
c_\fnum(\xi) \equiv 
\cos\left(\frac{|\xi|^2}{4\pi\fnum}\right)
\end{aligned}
 \label{eq:ctf}
\end{equation}
the corresponding transfer functions.
For a review on the validity conditions of the CTF approximation we refer to \cite{Turner2004}.

\subsection{CTF-based phase retrieval}

The task of phase retrieval is to reconstruct the unknown object $f$ from a set of intensity-only
holograms $I_j$, where $j = 1,...,J$, are holograms at different Fresnel numbers $\fnum_j$, e.~g. by
variation of the defocus distance. Using the linear CTF operator poses are linear inverse problem.
We express multiple holograms $I_j$ by using the linearity of the CTF in a matrix formalism that
we adapt from \cite{Kostenko2013}.

If we write the holograms as a stacked vector $I$, 
define the transfer matrix $M$, and 
(with a slight abuse of notation, identifying 
$\mathbb{C}$ with $\mathbb{R}^2$) 
express the complex-valued
object as vector $(i\phi + \mu) \simeq (\phi, \mu)^\top$, then 
\begin{equation}
 I =
  \begin{pmatrix}
   I_1 \\
   \vdots \\
   I_J
  \end{pmatrix},
 \qquad
 M = 2
 \begin{pmatrix}
  s_1 & c_1 \\
  \vdots & \vdots \\
  s_J & c_J
 \end{pmatrix},
  \qquad
 f =
  \begin{pmatrix}
   \phi \\
   \mu
  \end{pmatrix},
  \label{eq:ctf_vec}
\end{equation}
where $s_j= s_{\fnum_j}$, $c_j=c_{\fnum_j}$ denote the corresponding transfer functions defined in \eqref{eq:ctf} at Fresnel
number $\fnum_j$.
The matrix CTF operator is thus given by,
\begin{equation}
  I \simeq L(f) = 1 + \IFT M \FT(f),
\end{equation}
where the Fourier transforms $\FT$, $\IFT$ and scalar multiplication by two are meant element-wise.
Note that we use the diagonality of the Fresnel propagator in Fourier space and
thus $M$ is expressed \textit{per frequency}.

Phase retrieval amounts to the solution of the regularized linear inverse problem,

\begin{equation}
 f_\dagger = \argmin_f \left\{ \lnorm{ L(f) - I }^2 + \lnorm{A \FT f}^2 \right\},
 \label{eq:ctf_ip}
\end{equation}
where $A = A(\xi)= \diag\left(2\sqrt{\alpha_\phi(\xi)}, 2\sqrt{\alpha_\mu(\xi)}\right)$
denotes a frequency-weighted Tikhonov-type regularization operator, with $\alpha_\phi, \alpha_\mu \geq 0$.

The solution of least-squares problem in \eqref{eq:ctf_ip}
for $\alpha_\phi, \alpha_\mu > 0$ is given in closed-form by,
\begin{align}
 f_\dagger = \left( L^*L + \IFT A^*A \FT \right)\inv (L^* I)
  = \IFT\left[ \left( M^* M + A^*A \right)\inv \left(M^* \FT(I-1) \right) \right],
\end{align}
where the superscript $*$ denotes the adjoint. Here we used, that the Fourier transform is unitary, $\FT^* = \FT^{-1}$.
Explicitly, the solution in Fourier space is,
\begin{align}
    \begin{pmatrix}
		\FT{\phi}_\dagger \\
		\FT{\mu}_\dagger
	\end{pmatrix}
	=
	\frac{8}{D}
	\begin{pmatrix}
		\alpha_\mu + \sum_j c_j^2 & -\sum_j c_j s_j \\
		-\sum_j c_j s_j & \alpha_\phi + \sum_j s_j^2
	\end{pmatrix}
	\begin{pmatrix}
		\sum_j s_j (\FT I_j - 1) \\
        \sum_j c_j (\FT I_j - 1)
	\end{pmatrix},
 \label{eq:ctf_full_lstsq}
\end{align}
where $D = \det(M^*M + A^*A)$ (see appendix \ref{sec:determinant}). For readability we have omitted the summation limits
which are $j=1$ to $j=J$.
Written in
components, this coincides with the
formula presented in \cite{Zabler2005}. We remark  that due to \eqref{eq:ctf_full_lstsq} 
 phase retrieval without further a-priori information
is only uniquely feasible for \textit{at least two holograms}, i.e., $J \geq
2$. This goes along with the intuition that a meaningful estimation of two unknowns ($\phi$ and
$\mu$) from a single data point is not uniquely possible \textit{without additional assumptions on
the the solution}. As shown in \cite{maretzke:15,maretzke2017stability}, 
the assumption of compactness of the support 
of $\phi$ and $\mu$ suffices to restore uniqueness 
and even stability.

\subsubsection{Homogeneous objects}
\label{sec:homobject}


One ubiquitous assumption in XPCI is that the sample solely composed by a single material, what we
refer to 
as \textit{homogeneous} or \textit{single material object} (SMO)~\cite{Paganin2002}. This implies,
that phase and absorption images are proportional by $\frac{\mu}{\phi} = \frac{\beta}{\delta} =
\gamma = \mathrm{const}$, where $\delta$ 
and $\beta$ are defined after \eqref{eq:Fresnel_prop}. 
Thus, the number of unknowns effectively reduces from two to one, i.~e. $f = (i + \gamma)\phi$, but
introduces an additional parameter $\gamma$ that have to be determined \textit{a priori}.

Inserting the homogeneous object assumption into the CTF model, we get
\cite{Cloetens1999,Cloetens1999a}
\begin{equation}
  H_\gamma(\phi) = 1 + \IFT M_\gamma \FT\phi,
  \quad\mathrm{with}\quad
  M_\gamma = 2\begin{pmatrix}
              s_1 + \gamma c_1 \\
              \vdots \\
              s_J + \gamma c_J
             \end{pmatrix}.
  \label{eq:m_hom}
\end{equation}
Note that $M_\gamma$ is a $(J\times 1)$ matrix and the unknown variable is $\phi$. By the
reduction of the unknowns from two to one, also reconstructions from single-exposures are possible,
but technically only for homogeneous objects with known and well-defined $\gamma$.

Contrary to this, many samples are composed of different materials with significantly different
refractive indices. We call this kind of objects \textit{inhomogeneous} or multi-material objects.
A prominent example is contrast-enhancing staining. For inhomogeneous samples, the
homogeneous object assumption is an inexact model. A unique choice of $\gamma$ for such samples is
not possible.

\subsubsection{Constrained CTF}\label{sec:cctf}

Standard CTF-based phase retrieval in form of Fourier filters does not allow to impose prior
knowledge of the sample on the reconstruction. Object
constraints, such as a spatial support, (non-)negativity or box-constraints. Recently, we reported
on a framework, which allows to incorporate such constraints to CTF phase retrieval with
homogeneity assumption \cite{Huhn2022}.

Mathematically, such object priors restrict the space of possible solutions to a proper 
closed, convex subset $C$,
\textit{i.~e.} we require $f_\dagger \in C \subsetneq L^2(\RR^2)$. Thus, constrained CTF phase
retrieval corresponds to the minimization of
\begin{equation}
 f_\dagger = \argmin_{f \in C} \left\{ \lnorm{ L(f) - I }^2 + \lnorm{A \FT f}^2 \right\}.
 \label{eq:cctf_ip}
\end{equation}

For \eqref{eq:cctf_ip} no closed-form or single step solution exists. However, a solution can 
be computed iteratively with the Alternating Direction Method of Multipliers (ADMM)
\cite{Boyd2011,Beck2017}.
%
The iteration update for CTF-ADMM is given by,
\begin{subequations}
 \begin{align}
  f_{k+1} &= \argmin_{f \in L^2(\RR^2)} \left\{ \lnorm{ L(f) - I }^2 +
        \lnorm{A \FT f}^2 + \tau\lnorm{f - {(g_k - h_k)}}^2 \right\} \nonumber \\
    &= \left( L^*L + \IFT A^*A\FT +\tau\eye \right)\inv (L^* I + \tau {(g_k - h_k)})
\label{eq:prox_ctf} \\
 g_{k+1} &= \Pi_C (f_{k+1} + h_k) \label{eq:prox_constraints} \\
 h_{k+1} &= h_k + f_{k+1} - g_{k+1},
 \end{align}
 \label{eq:ctf_admm}
\end{subequations} 
where $\eye$ is the identity matrix, $g, h$ are auxiliary variables initialized with $g_0, h_0 = 0$, $\tau > 0$ is a proximity
parameter and $\Pi_C(f):=\mathrm{argmin}_{h\in C}\|f-h\|$ is the metric projection onto $C$. In \eqref{eq:ctf_admm} we see the
splitting into the so-called proximal mapping
\cite{Bauschke2017,Combettes2011} of the CTF inversion \eqref{eq:prox_ctf} and object constraints
\eqref{eq:prox_constraints}. Note that both solutions are given in closed form and no
sub-iterations are needed. We apply a Nesterov-type accelerated ADMM variant; for its details we
refer to \cite{Goldstein2014}.
\\ \\
To summarize, we now have four possible combinations of CTF phase retrieval variants,
\textit{i.~e.}
\begin{enumerate}
 \item with homogeneous object assumption, without object constraints (HomCTF),
 \item with homogeneous object assumption, with object constraints (HomCTF $\phi \in C$),
 \item without homogeneous object assumption, without object constraints (CTF),
 \item without homogeneous object assumption, with object constraints (CTF $f \in C$).
\end{enumerate}

\section{Stability analysis}\label{sec:stability}

In this section we analyze the influence of the homogeneity assumption on the stability of the
phase retrieval and consequences for phase retrieval without it. To do so, we compare the singular
values of the corresponding transfer matrices, which can be interpreted as effective contrast
transfer or signal per frequency. The higher the singular value, the higher the effective signal
and thus the more robust is the phase retrieval against noise. Vice versa, near-zero or zero
singular values correspond to vanishing effective signal, and near singular matrices, whose
inversion for phase retrieval could be heavily distorted by noise.

We note that the singular
values of the not regularized and unconstrained forward operator are overly pessimistic
stability predictions. The use of additional a-priori assumption such as object constraints,
e.g. compact spatial support or (non-)negativity, has strong influence
\cite{maretzke2017stability}. Nevertheless, the singular values give an access to the general
behavior of the (linearized) system and could be used to optimize experimental parameter, for
example an optimal set of Fresnel numbers as reported for the homogeneous CTF in \cite{Zabler2005}.

Here we use the normalized transfer matrix $\tilde{M}$, where we omit the multiplication by
two as defined in \eqref{eq:ctf_vec}; likewise, for the homogeneous transfer matrix $\tilde{M}_\gamma$.
Thus, direct computation of the (smallest) singular value $\sigma_-$, of the $\tilde{M}$, yields,
\begin{equation}
 \sigma_-^2(\tilde{M})(\xi) = \frac{J}{2} - \frac{1}{2}
  \left(
   J^2 - 4\sum_{k = 1}^J \sum_{j=k+1}^J \sin^2\chi_{kj}(\xi)
  \right)^{\frac{1}{2}},
 \label{eq:smin}
\end{equation}
where we used the difference phase chirp $\chi_{kj}$, given by
$
 \chi_{jk}(\xi) \equiv \frac{|\xi|^2}{4\pi} \left( \frac{1}{\fnum_k} - \frac{1}{\fnum_j} \right)
  \equiv \frac{|\xi|^2}{4\pi \fnum_{kj}},
$
with \textit{difference Fresnel number} $\fnum_{kj}^{-1} \equiv \fnum_{k}^{-1} - \fnum_{j}^{-1}$, 
introduced in \cite{maretzke2017stability}.
In the following we will omit the
explicit $\xi$ dependence.
We note that the summation in \eqref{eq:smin} has $N(J) = \frac{J^2 - J}{2}$ terms, \textit{e.~g.}
$N(2) = 1$ or $N(3) = 3$. Especially, for a single hologram, $J=1$, no summand remains and
$\sigma_-(\tilde{M}) = 0$ for all $\xi$. Thus, phase retrieval from a single hologram without homogeneity
assumption is instable for all frequencies, without additional regularity assumptions.

\begin{figure}
 \centering
 \includegraphics{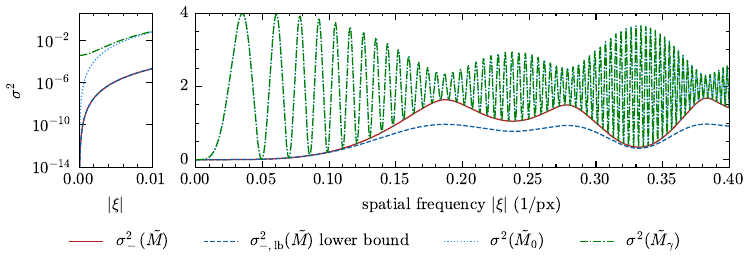}
 \caption{Squared smallest singular values $\sigma$ of CTF Fourier transfer matrices $\tilde{M}$ without
homogeneity assumption respectively $\tilde{M}_\gamma$ ($\gamma = 0.01$) with homogeneity for a set of four Fresnel numbers
 $\fnum_j$ \numlist{2.50e-03; 2.45e-03; 2.40e-03; 2.39e-03}.
\textbf{Left} subfigure shows zoom into lower frequencies behavior, \textbf{right}
larger frequency range. We can observe that $\sigma_-^2(M)$ is a lower envelope for
$\sigma^2(\tilde{M}_0)$. Note that the oscillations for $\sigma_-^2(\tilde{M})$ depend on the
\textit{difference} Fresnel numbers instead of the Fresnel number, as $\sigma(\tilde{M}_\gamma), \sigma(\tilde{M}_0)$ do.
An approximation from below for $\sigma_-^2(\tilde{M})$ is shown by $\sigma_{-,\mathrm{lb}}^2(\tilde{M})$  defined in
\eqref{eq:smin_lower}, with good agreement if $\sigma^2 \ll 1$.}
 \label{fig:stability}
\end{figure}

As we mention above, vanishing singular values correspond to instable phase retrieval for those
frequencies. Thus, we approximate $\sigma_-(\tilde{M})$ from below by
$\sigma_{-,\mathrm{lb}}(\tilde{M})$,
\begin{equation}
 \sigma_-^2(\tilde{M}) \geq \sigma_{-,\mathrm{lb}}^2(\tilde{M}) \equiv \frac{1}{J} 
 \sum_{k = 1}^J \sum_{j=k+1}^J \sin^2\chi_{kj},
 \label{eq:smin_lower}
\end{equation}
which shows good agreement for low contrast transfer $\sigma_-^2(\tilde{M}) \ll 1$, see Figure
\ref{fig:stability}.

More importantly, this allows us to make a direct comparison to the singular value of the
homogeneous CTF matrix $\tilde{M}_\gamma$, given by,
\begin{equation}
 \sigma^2(\tilde{M}_\gamma) = \sum_j (\sin\chi_j + \gamma \cos\chi_j)^2,
 \label{eq:smin_hom}
\end{equation}
with phase chirp $\chi_j \equiv \frac{|\xi|^2}{4\pi \fnum_j}$.

By comparing \eqref{eq:smin_hom} with \eqref{eq:smin_lower} it is evident that the lower stability
bounds of the inhomogeneous CTF has a similar structure as a pure phase CTF, \textit{i.~e.}
$\sigma^2(\tilde{M}_0) = \sum_j \sin^2\chi_j$, but with dependence on the \textit{difference} Fresnel number.

In fact, for a reconstruction of two holograms at different Fresnel numbers, the
stability of the inhomogeneous CTF can be seen as proportional to a \textit{single hologram} pure
phase CTF reconstruction, but at effective difference Fresnel number \cite{maretzke2017stability}.
Therefrom, we argue that inhomogeneous CTF reconstructions \textit{without additional constraints}
need at least two holograms at different Fresnel numbers. To stabilize the reconstruction, three or
more holograms at different Fresnel numbers are favorable.

The Fresnel holographic phase retrieval is known to be sensitive to noise in the lower frequencies,
especially in case of a pure phase object \cite{Langer2021IEEE,Zabler2005}, which relates to the
vanishing singular value for $\xi \to 0$. In case of a homogeneous, non-pure phase object,
\textit{i.~e.} $\gamma > 0$, this is circumvented by the contribution of the non-zero cosine term,
see Fig. \ref{fig:stability} especially the left sub-figure. Thus, the homogeneity assumption has an important stabilizing effect.

In Fig. \ref{fig:stability} we illustrate our above analysis by an plot of the (squared) singular values for an exemplary set
of Fresnel numbers of $\fnum_j$, which are \numlist{2.50e-03; 2.45e-03; 2.40e-03; 2.39e-03}, and set 
$\gamma = 0.01$.
The corresponding six difference Fresnel numbers $\fnum_{kj}$ are 
\numlist{4.41e-01; 1.23e-01; 1.18e-01; 9.28e-02; 6.00e-02; 5.28e-02}. In Fig. \ref{fig:stability} we see, that the inhomogeneous CTF (smallest)
singular value $\sigma_-^2(\tilde{M})$ is a lower envelope of the pure phase CTF singular value $\sigma_-^2(\tilde{M}_0)$.

\subsection{Regularization strategy}

Based on our observation that the stability of the inhomogeneous CTF relates to the pure phase
CTF's stability at effective difference Fresnel numbers, we conclude that a similar regularization
strategy as usually employed for the homogeneous CTF should be suitable. \textit{I.~e.} a
frequency-dependent (or frequency weighting) regularization value $\alpha(\xi)$ that separates the
manifested singularity around $\xi \approx 0$ and higher frequencies regime with transition at the
first pure-phase CTF maxima \cite{Cloetens1999,Huhn2022}.

To account for the different orders of magnitudes of phase and absorption contrast for weak samples 
(for $\sim\qty{10}{\keV}$ and light elements or compounds, we have $\delta \sim \num{e-6}$ and $\beta \sim
\num{e-9}$), we employ 
regularization in phase and absorption individually. Hence, the
Tikhonov-type frequency regularization operator $A$ becomes a diagonal operator given by 
$A(\xi) = \diag\left(2 \sqrt{\alpha_\phi(\xi)}, 2 \sqrt{\alpha_\mu(\xi)}\right)$.

\section{Experimental evaluation}
%
%

\subsection{Experimental data}

We evaluate the performance of CTF-based phase retrieval with experimental data taken on the
so-called \textit{Göttinger Instrument for Nano-Imaging with X-rays} (GINIX)
\cite{Salditt_JoSR_2015} located at the beamline P10 of the PETRA III storage ring at DESY, Hamburg,
Germany. We used the waveguide focussed high-resolution nano-imaging setup in magnifying divergent
beam geometry\mynote{cite}. The X-ray waveguide acts as quasi-point source for a clean illumination
wavefront. For the experiment the X-ray energy was set to $\qty{13.8}{\keV}$. The
diffraction patterns were recorded with the scintillator-based Andor Zyla 5.5 sCMOS camera, with a
pixel size of $\qty{6.5}{\um}$.

\begin{figure}
 \centering
 \includegraphics{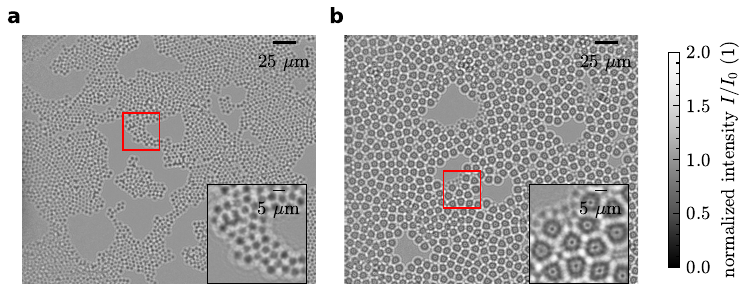}
 \caption{Exemplary flat-field corrected holograms of colloidal sample used for phase retrieval.
The effective Fresnel numbers is $\fnum = \num{1.835e-03}$ at an effective pixel size
$\qty{127.2}{\nm}$.
Subfigure \textbf{a} shows mixture of \qty{4.27}{\um} silica and \qty{4.24}{\um} polystyrene, \textbf{b}
with \qty{9.36}{\um} silica and \qty{4.24}{\um} polystyrene. Insets corresponds to insets in reconstructions below.}
 \label{fig:colloids_holograms}
\end{figure}

By transaxial translation of the sample along the optical axis different propagation distances
between sample and detector, thus Fresnel numbers, are recorded. By the Fresnel scaling
theorem (see \textit{e.~g.} \cite{Paganin2006,Salditt_Nano_2020}) the near-field diffraction
patterns taken in the divergent beam geometry (spherical wavefronts) can be transformed into
an equivalent effective parallel beam coordinate system (plane wavefronts)\mynote{cite}.
In line with our results from \autoref{sec:stability}, we acquired holograms at four defocussed
distances corresponding to effective (parallel beam) Fresnel numbers $\fnum_j$ 
\numlist{1.84e-03; 1.81e-03; 1.78e-03; 1.73e-03}. To increase the signal-to-noise
ratio, each hologram is an average of 100 images. Intensity images are normalized by principal
component analysis (PCA) synthesized empty images \cite{Hagemann2021XFEL}. Examples of the
holograms are shown in \autoref{fig:colloids_holograms}.

To violate the homogeneous object assumption, we prepared samples with two materials, with
different but known refractive properties. To this end, we mixed silica (SiO$_2$) and polystyrene
(PS) colloids. One sample with similar diameters of \qty{4.27}{\um} silica and \qty{4.24}{\um} polystyrene
(see
Fig. \ref{fig:colloids_holograms}\textbf{a}), the other with \qty{9.36}{\um} silica and \qty{4.24}{\um}
polystyrene (see Fig. \ref{fig:colloids_holograms}\textbf{b}). To corresponding refractive
properties at $\qty{13.8}{\keV}$ are given in \autoref{tab:ref_params}.

\begin{table}
 \centering
  \begin{tabular}{l c c c c}
   Compound & $\delta$& $\beta$& $\gamma = \frac{\beta}{\delta}$& density (\unit{\g\per\cm^3}) \\ \hline \hline
    Silica (SiO$_2$) & \num{1.86e-06} & \num{1.60e-08} & \num{8.61e-03} & 1.7 \\ \hline
    Polystyrene (PS) & \num{1.23e-06} & \num{7.08e-10} & \num{5.75e-04} & 1.1 \\ \hline
  \end{tabular}
  \caption{Real $\delta$ (phase shift) and imaginary $\beta$ (attenuation) parts of the complex
refractive index for listed compounds at \qty{13.8}{\keV} X-ray energy. Source:
\cite{Schoonjans2011}.}
  \label{tab:ref_params}
\end{table}

\subsubsection{Silica and polystyrene colloids of equal size}
\label{sec:colloids14}

We begin the experimental validation with the equally sized colloids mixture of silica and
polystyrene. By this, the primary difference between the colloids is expected to the be different
ratios between absorption and phase shift index, $\gamma = \beta / \delta$.

\begin{figure}
 \centering
 \includegraphics{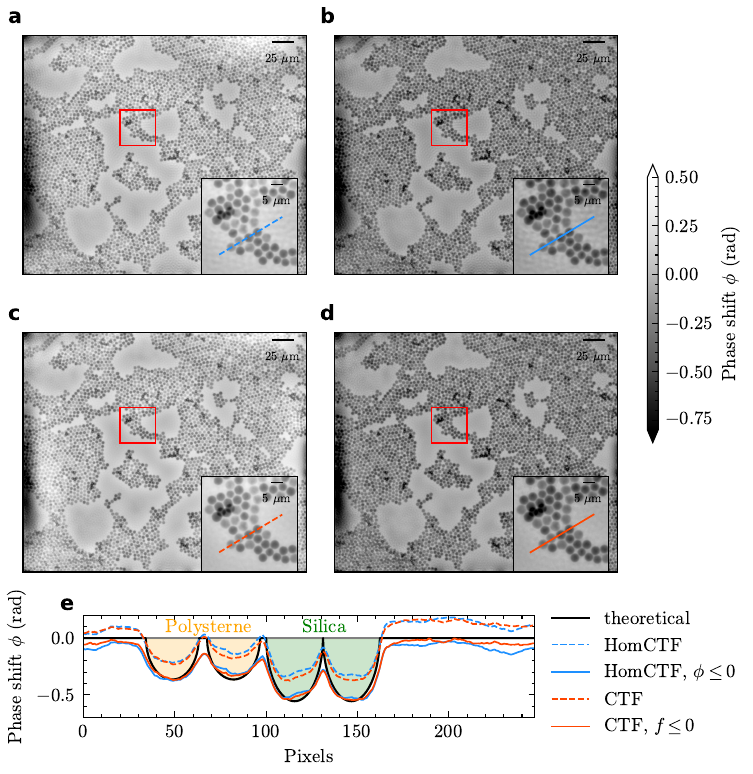}
 \caption{CTF phase reconstructions of \qty{4.27}{\um} silica and \qty{4.24}{\um} polystyrene colloids
sample. Top row \textbf{a, b} shows standard CTF-based phase retrieval with homogeneous object
assumption, denoted as \textit{HomCTF}. Bottom row \textbf{c, d} show our proposed approach without
homogeneous object assumption.
Left column \textbf{a, c} reconstructions are single-step without object constraints; right column
are reconstruction with negativity constraint ($f \leq 0$) computed with ADMM. Overall,
all reconstructions agree. This shows, faithful CTF reconstructions are possible also without the
homogeneity assumption. Moreover, CTF inversion without homogeneity constraint show less
background oscillations; compare top row to bottom row. Hence, the background with homogeneity
constraint is more uniform. Subfigure \textbf{e} shows a line profile indicated in the
inset zoom plots of the reconstructions is shown over a theoretical expectation of the phase
shifts. Pixel size is \qty{127.2}{\nm}.}
 \label{fig:colloids14}
\end{figure}

We compared the four combinations of CTF phase retrieval, listed in \ref{sec:cctf}. The
retrieved phase images are shown in \autoref{fig:colloids14}. The reconstruction parameters are
listed in \autoref{tab:ctf_params}. To compare the influence of the homogeneity assumption we need
to compare top-to-bottom; comparison between unconstrained and negativity constrained
reconstruction is left-to-right.

\begin{table}
 \centering
  \begin{tabular}{l c c}
   \hline
    $\alpha_\gamma(\mathrm{low; high})$ & $\num{0}$ & $\num{5e-3}$ \\ \hline
    $\gamma$ = $\gamma(\mathrm{SiO}_2)$ & \num{8.61e-3} & -- \\ \hline
    $\tau$ & \num{1e-2} & -- \\ \hline
    \hline
    $\alpha_\phi(\mathrm{low; high})$ & $\num{6e-5}$ & $\num{5e-3}$ \\ \hline
    $\alpha_\mu(\mathrm{low; high})$ & $\num{0}$ & $\num{5e-1}$ \\ \hline
    $\tau$ & \num{1e-4} & -- \\ \hline
  \end{tabular}
  \caption{Regularization and reconstruction parameters for homogeneous (first three rows) and inhomogeneous (last
three rows) CTF. Two-level frequency regularization $\alpha(\xi)$ is applied with cut-off frequency
at the first pure phase CTF maxima. For the homogeneity regularization the $\beta/\delta$-ratio of
the stronger absorbing material (here silica) is used, which produces less artifacts. $\tau$
denotes to proximity parameter for the constrained CTF-ADMM, see \eqref{eq:ctf_admm}.}
  \label{tab:ctf_params}
\end{table}

Overall, all phase reconstructions agree. If we look more closely, we observe oscillatory
artifacts spreading across the image in the CTF reconstructions with homogeneity assumption (see
\autoref{fig:colloids14}\textbf{a-b}). These oscillation can also be seen in the line profile given
in subfigure \autoref{fig:colloids14}\textbf{e}. In direct comparison these are significantly
reduced in the inhomogeneous CTF reconstruction.
Albeit, these can also be observed at the edges of the inhomogeneous
reconstructions. We attribute this to the 2D Gaussian-shaped illumination, which was focused centrally
in the images. Hence, closer to the edges of the detector the illumination was lower and in turn
the noise ratio higher. Due to the lower stability of the inhomogeneous CTF, see
\autoref{sec:stability}, this has a stronger influence on the reconstructions.

If we compare the negativity constrained reconstructions \autoref{fig:colloids14}\textbf{b,d} with
the unconstrained ones \autoref{fig:colloids14}\textbf{a,c}, we see that wrong positive phase
shifts (light gray areas) are corrected. Furthermore, low frequency background variations are
suppressed. Furthermore, in the line profile \autoref{fig:colloids14}\textbf{e} we see that the
unconstrained solutions (dashed lines) qualitatively follow the theoretical line, but the
constrained solutions (solid lines) agree quantitatively.

\subsubsection{Silica and polystyrene with different diameter}
\label{sec:colloids11}

For a second sample we replaced the equal sized silica colloids with silica colloids of
\qty{9.36}{\um} diameter. Thus, the projected phase shifts for the silica colloids exceed
$\pi/4~\unit{\radian}$.

Generally, the result discussed in section \ref{sec:colloids14} are also confirmed for this sample.
In particular, here the oscillatory artifacts are pronounced for the homogeneous CTF and still
significantly suppressed for the inhomogeneous CTF. However, deviation from the expected phase
shift can be observed in both reconstructions. Most notably, the central ``dot'' within the
colloids is a persistent artifact in both reconstructions (and not a feature of the colloids).

\begin{figure}
 \centering
 \includegraphics{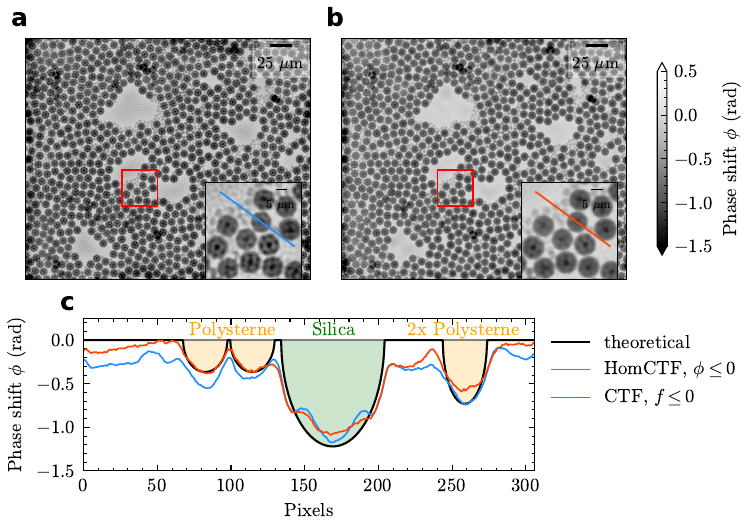}
 \caption{CTF phase reconstructions of \qty{9.36}{\um} silica and \qty{4.24}{\um} polystyrene
colloids sample. Subfigure \textbf{a} shows reconstruction with homogeneity assumption (HomCTF) and
\textbf{b} without. Both reconstructions are with negativity constraint, \textit{i.~e.}
$\phi$ respectively $f \leq 0$. Corresponding line scans as indicated in the phase images are shown
in \textbf{c}. The reconstruction with homogeneity shows prominent oscillatory artifacts, which are
present in the background as well as in the colloids, whereas the CTF without homogeneity overall
is closer to the theoretical curve, albeit deviations can be observed. The central dots within the
larger colloids are artifacts appearing in both reconstructions. Pixel size is \qty{127.2}{\nm}.}
 \label{fig:colloids11}
\end{figure}

\section{Discussion \& Outlook}

To summarize, we have compared CTF-based phase retrieval, with the homogeneous object
assumption (\textit{homogeneous CTF}), introduced in \autoref{sec:homobject}, and without it
(\textit{inhomogeneous CTF}). In combination with frequency-dependent regularization
individually in phase and absorption by an diagonal regularization operator $A =
\diag\left(2\sqrt{\alpha_\phi}, 2\sqrt{\alpha_\mu}\right)$ and objects constraints, 
for example (non-)negativity, imposed on
the reconstruction, we showed that inhomogeneous CTF inversion is possible, and even more so shows
less oscillatory artifacts compared to the homogeneous CTF.

For both datasets tested, the inhomogeneous CTF showed superior image quality and more faithful results
compared to the homogeneous CTF, especially when combined with negativity object constraint. In
contrast, if the same
dataset is reconstructed with the homogeneous CTF, we observe oscillatory artifacts,
see \autoref{fig:colloids14}\textbf{a,c} and \autoref{fig:colloids11}\textbf{a}. Thus, we
interpret the homogeneous CTF to be overly restrictive or inexact for non-homogeneous samples. As a
result, such reconstructions are imprinted with artifacts.

For the colloids with $\qty{4}{\um}$ diameter, as discussed in \autoref{sec:colloids14}, the
theoretical expected phase shifts are $\phi \leq \pi/4~\unit{\radian}$. In comparison to the sample
of \autoref{sec:colloids11} with larger colloids, the expected phase shifts partly exceed 
$\pi/4~\unit{\radian}$. For the latter, more pronounced oscillatory artifacts can be observed when
reconstructed with homogeneity assumption. Again, the inhomogeneous CTF is less imprinted with
those artifacts, albeit for the second sample deviations for larger phase shifts ($\geq
\qty{0.5}{\radian}$) can be seen.
Thus, we relate these artifacts in the homogeneous CTF to be party caused by nonlinearity errors, which
in turn can be better compensated in the less restricted inhomogeneous CTF, allowing qualitatively
superior results.

We note that the CTF inversion as linear inverse problem allows computationally fast
reconstructions, also with constraints imposed via ADMM \cite{Huhn2022}. In the unconstrained case,
it still is a one-step Fourier filter. The computational
complexity from the homogeneous to inhomogeneous CTF is doubled, since the unknown variables are
doubled from only $\phi$ to $f = (\phi, \mu)^\top$. Still, it has a considerably less
computational burden as any nonlinear method. This can potentially be utilized for qualitative
online reconstructions, directly on data acquisition.

Furthermore, we have analyzed the influence on the stability of the homogeneity assumption and
found the stability of the inhomogeneous CTF to be the lower envelope of the stability of the
pure phase CTF. We approximated the frequencies with near zero singular values, where inversion
is sensitive to noise, of the inhomogeneous CTF by a structurally similar term as the pure phase
CTF singular value. We conclude that the stability of the inhomogeneous CTF can be related to be
lower bounded by the pure phase CTF singular value, but at effective \textit{difference Fresnel
numbers}.

As a result, the well-known low-frequency instability of the (pure phase) CTF inversion also
applies to the inhomogeneous CTF even more, but depending on the difference Fresnel number. Thus, optimal
experimental settings are governed by the difference Fresnel number. Furthermore, we showed by
example of four holograms recorded at different Fresnel numbers, that an inhomogeneous CTF reconstruction is
feasible. The addition of prior knowledge via a constrained optimization using the ADMM algorithm
yield results in good agreement with the theoretical predictions, see Fig.
\ref{fig:colloids14}\textbf{e} respectively \ref{fig:colloids11}\textbf{c}.

Moreover, as we see for the larger colloids, in sec. \ref{sec:colloids11}, nonlinear effects start
to interfere with linearized reconstruction methods. Also, stronger violations of the homogeneity
assumption involve higher nonlinearity. Hence, naturally the question of the impact of the
homogeneity constraint and its omission on nonlinear reconstruction raises. Albeit inhomogeneous
nonlinear solvers exists, these usually need many iterations to converge.

A possible nonlinear extension to the CTF is a so-called \textit{frozen Gauss-Newton} iteration
\cite{Jin2010}. Here, in contrast to a regular (iteratively regularized) Gauss-Newton method, the
Fréchet derivative of the Fresnel imaging operator \eqref{eq:fresnel_imag} is not (computationally
expensive) computed in each iteration, but kept at the initial point $f = 0$ (weak object
limit). The derivative than coincides with the CTF operator \cite{Maretzke2016}. This can be seen
as addition of a nonlinearty term to the linear inverse problem of the CTF defined in
\eqref{eq:ctf_ip}.

Furthermore, in view of the stabilizing effect of the homogeneity constraint, as we discussed in
\autoref{sec:stability}, a reasonable regularization strategy could be to regularize the deviation
from the homogeneous solution instead in the basis of phase and absorption. Finally, depending on the object and experimental case, our work can also be used to justify the current use of the homogeneous CTF, based on 
dissecting the effects of (weak) non-homogeniety and non-linearity.  


\begin{backmatter}
\bmsection{Funding} Deutsche Forschungsgemeinschaft (DFG) - Project number 432680300 (SFB 1456-C03).

\bmsection{Disclosures} The authors declare no conflicts of interest.

\bmsection{Acknowledgments}
We thank Simon Huhn for the inspiring discussions and helpful comments.

\bmsection{Data availability} Data underlying the results presented in this paper are not publicly
available at this time but may be obtained from the authors upon reasonable request.

\end{backmatter}

\bibliography{ctf}

\appendix
\section{Determinant of the transfer matrix}
\label{sec:determinant}
The determinant $D$ used in \eqref{eq:ctf_full_lstsq} with regularization 
$A = \diag\left(2\sqrt{\alpha_\phi}, 2\sqrt{\alpha_\mu}\right)$ is given by 
\begin{equation}
\begin{aligned}
 D &\equiv \det(M^*M + A^*A) \\ 
  &= \det\left\{ 
  \begin{pmatrix}
      4 \sum_j s_j & 4 \sum_j s_j c_i \\
      4 \sum_j s_j c_i & 4 \sum_j s_j
  \end{pmatrix}
  +
  \begin{pmatrix}
      4 \alpha_\phi & 0 \\
      0 & 4 \alpha_\mu
  \end{pmatrix}
 \right \} \\
  &=
  \left( 4\alpha_\phi + 4\sum_j s_j^2 \right) \left( 4\alpha_\mu + 4\sum_j c_j^2 \right)
  - \left( 4\sum_j c_j s_j \right)^2 \\
  &= 16 \left\{
  \alpha_\mu\alpha_\phi + \alpha_\mu\sum_j s_j^2 + \alpha_\phi\sum_j c_j^2 + \left(\sum_j s_j^2\right) 
  \left( \sum_j c_j^2\right) - \left( \sum_j c_j s_j \right)^2 \right\},
 \end{aligned}
\end{equation}
where sums are meant as summation from $j=1$ to $j=J$.

\end{document}